\documentclass[preprint]{revtex4-1}
\usepackage{amsmath,amssymb}

\renewcommand{\P}{{\mathbb P}}

\newcommand{\Z}{{\mathbb Z}}
\newcommand{\tr}{{\rm tr\,}}

\begin{document}

\title{What is the minimal supersymmetric Standard Model  \\ from F-theory?}
\author{Kang-Sin Choi}
\email{kangsin@ewha.ac.kr}
\affiliation{Scranton College, Ewha Womans University, Seoul 120-750, Korea, \\
Korea Institute for Advanced Study, Seoul 130-722, Korea}

\begin{abstract}
We construct gauge theory of $SU(3)\times SU(2)\times U(1)$ by spectral cover from F-theory and ask how the Standard Model is extended under minimal assumptions on Higgs sector. For the requirement on {\em different} numbers between Higgs pairs and matter generations (respectively one and three) distinguished by $R$-parity, we choose a universal $G$-flux obeying $SO(10)$ but slightly breaking $E_6$ unification relation. This condition {\em forces} distinction between up and down Higgs fields, suppression of proton decay operators up to dimension five, and existence and dynamics of a singlet related to $\mu$-parameter.
\end{abstract}
\maketitle

\section{Introduction}

We explore a supersymmetric extension of the Standard Model (SM) from F-theory, under certain minimal assumptions on Higgs sector. Construction from 
F-theory, admitting dual $E_8 \times E_8$ heterotic string, naturally yields a realistic Grand Unified Theory (GUT) of gauge group along $E_n$ series, including SM itself \cite{GUT,Enunif}. In string derived models, however, such unification relation is so strong that it has been very difficult to understand the nature of Higgs doublet in this context, namely how to embed it to a larger GUT representation and why its observed number should be different from that of quark and lepton generations. The main result of this paper is that F-theory can control such features, implying some nontrivial phenomenological consequences. For example it gives us understanding on how can we distinguish up and down-type Higgs fields and what are  the properties of the $\mu$-parameter in the Minimal Supersymmetric Standard Model (MSSM).

We first build $SU(3) \times SU(2) \times U(1)_Y$ gauge group, without aid of an intermediate Grand Unification. 
By specifying a spectral cover of structure group $S[U(5)\times U(1)_Y]$, its commutant  in $E_8$ survives as the SM gauge group \cite{CK,Choi:2010nf}. The spectral cover is a systematic way to construct (poly)stable vector bundle in dual heterotic string, if the compact manifold admits elliptic fibration with a (usually called zero) section \cite{FMW}.  Although the desired spectral cover is obtained by tuning parameters of an $SU(6)$ cover \cite{TW,Blumenhagen:2006ux,Marsano:2009gv}, the existence of the $U(1)_Y$ gauge group is not guaranteed until the following two requirements are met. First, elliptic fiber of heterotic string admits more {\em global} section(s) than the zero section, since monodromy should not mix the single cover for $U(1)_Y$ from extension to non-abelian structure group \cite{Grimm:2010ez,Katz:2011qp}. This we do by tuning elliptic fiber as well \cite{CH}. Second, the corresponding gauge boson should not acquire mass by St\"uckelberg mechanism, which we evade by {\em not} turning on $G$-flux along this direction.

To obtain chiral spectrum in four dimension, we also have to turn on so-called $G$-flux \cite{DW3,Hayashi:2008ba}. It is important to note the unique feature of F-theory that the {\em unbroken group} is solely determined by the spectral cover and $G$-flux only affects the number of {\em zero modes}. Thus, as long as the spectral cover has the structure group $S[U(5)\times U(1)_Y]$ the unbroken group is the SM group. To distinguish Higgs from lepton doublet in supersymmetric model by $R$-parity and also to impose different number of Higgs from that of the unified matter multiplets, it will turn out that the structure group of spectral cover is singled out to be $S[U(3)_\perp \times U(1)\times U(1)\times U(1)]$.

If the structure group is semi-simple and possibly plus abelian, we can {\em partly} turn on the $G$-flux on a subgroup. For example, turning on $G$-flux on $SU(3)_\perp$ group, the resulting number of generation obeys the unification relation of the commutant group $E_6$, predicting the same number of fields belonging to $\bf 27$ multiplet of $E_6$, thus the number of Higgs doublet should be the same as that of quark generations. This relation can be relaxed, on the other hand, if $G$-flux is on $SU(4)$, giving unification relation of $SO(10)$. This is attempted in the previous work \cite{CK}, but the number of Higgs doublet is also totally determined to be undesirable one.
To control them {\em differently} we turn on two different $G$-fluxes along its subgroup $S[U(3)_\perp \times U(1)] \subset SU(4)$ with one more free parameter.
Since it does not obey $E_6$ unification the number of Higgs pair can be different to that of matter quarks, to be three and one, respectively, adjusted by $U(1)$ flux strength. The entire flux still does not touch $SO(10)$ direction, the model possesses $SO(10)$ unification relation thus we have the same number of quarks and leptons, as well as that of right-hand neutrinos.

Finally, the four dimensional interactions follow from gauge invariant terms of the higher dimensional effective Lagrangian by dimensional reduction \cite{BHV,Grimm:2010ks}. The invariance under the various $U(1)$ groups from the above spectral cover plays the role of selection rule. The structure of these symmetries predicts aforementioned phenomenological features. Also, we analyze the vacuum configuration giving proper interactions evading nucleon decays.

\section{Gauge group}

The model is obtained from F-theory compactification on elliptic Calabi--Yau fourfold with a section, admitting heterotic dual. 
The dual heterotic string is compactified on elliptic Calabi--Yau threefold $Z \to B_2$ with a section, which is usually called as the zero section. To have a globally well-defined $U(1)$ used by the SM gauge group and its constructing spectral cover, we need {\em another global section} than the zero section on the fiber to parameterize the dual point to the line bundle of the $U(1)$ structure group \cite{CH}. Globally, this point will not be mixed by monodromy with other points parameterizing other spectral covers, as we move around the entire base $B_2$. Let the canonical bundle of the base $B_2$ be $K_{B_2}$. We choose the coordinate of such point as $(x_1,y_1)$, which are global holomorphic sections $x \in \Gamma(B_2,O(K_{B_2}^{-2}))$ and $y \in\Gamma(B_2,O(K_{B_2}^{-3}))$, and the elliptic equation has a form
\begin{equation} \label{moresection}
 (y-y_1)(y+y_1) = (x-x_1)(x^2+x_1 x+x_1^2 + f)
\end{equation} 
where $f \in \Gamma(B_2,O(K_{B_2}^{-4}))$. 

We construct the spectral cover for the structure group $S[U(5)\times U(1)_Y]$ as follows \cite{CH,CK,Choi:2010nf},
\begin{equation} \label{su5u1cover} 
 a_0 + a_2 x + a_3 y + a_4 x^2 + a_5 xy + a_6 x^3 = 0, 
 \end{equation}
with tuning of parameters
\begin{equation}\begin{split} 
 a_0 &= d_0, \ a_2=d_2 + b_1 d_1, \ a_3= d_3+b_1 d_2, \\
 a_4 &= d_4 + b_1 d_3, \ a_5 = d_5 + b_1 d_4,\ a_6= b_1 d_5, \\ 
\end{split}
\end{equation}
with the constraint $ d_1 + b_1 d_0 = 0$. Here $a_m \in \Gamma(B_2,O(K_{B_2}^{m}))$ are globally defined and no approximation, e.g. of Higgs bundle type is used.
 In addition, to guarantee the existence of a global section with holomorphic parameters, we need further factorization condition \cite{CH}
\begin{equation}
 f  = b_1^2 F, \  g= -b_0^2 F, \ d_0=b_0 d, \ d_1 = -b_1 d,
\end{equation}
where the topological properties of $d$ and $F$ can be deduced from those of $f,g,b_0,b_1$.
Since the global section $(x_1,y_1)=(b_0^2/b_1^2, \pm b_0^3/b_1^3)$ is on this spectral cover (\ref{su5u1cover}), the coordinate values will be expressed in terms of the parameters $b_1$ and $d_m$.
We can take an analogy of Higgs bundle for large $x$ and $y$ to plug well-known solution so far (but we do not stick to it since our description is valid for all $x$ and $y$ as long as the stable degeneration limit is valid): Each coefficient $d_m$, parameterizing the positions of the covers, is related to the elementary symmetric polynomial of degree $m$, out of weights of the fundamental representations ${\bf 5}_1+{\bf 1}_{-5}$ of the $S[U(5)\times U(1)_Y]$. 
The surviving group on $B_2$ is the commutant, the SM group $SU(3) \times SU(2) \times U(1)_Y$. This a sufficient specification, so that it provides the information on the unbroken gauge group \cite{Beetal}.

In the stable degeneration limit \cite{FMW,Hayashi:2008ba}, we can convert the equations (\ref{moresection}) and (\ref{su5u1cover}) into the singularity equation corresponding to the SM group
\begin{equation} \label{weieq} \begin{split}
y^2 & = x^3 + (d_5 + d_4 b_1) xy + (d_3 + d_2 b_1)(b_1 d_5+ z)y z \\
& +(d_4 + d_3 b_1) x^2z + (d_2 - b_1^2 d)(b_1 d_5 +  z)x z^2   + d (b_1 d_5 +  z )^2 z^3  \\
 &+ b_1^2 F x z^4 - F z^6  ,
  \end{split}
\end{equation}
where $x,y$ are affine coordinates of $\P^2$ and $z$ is the coordinate of blown-up $\P^1$ in the stable degeneration \cite{Hayashi:2008ba}. Roughly, $z$ is a normal coordinate to $B_2=\{z=0\}$ inside the base of elliptic fibration $B$ in the F-theory side.
At the discriminant locus of (\ref{weieq}), we have the the SM gauge group \cite{Choi:2010nf}. Referring to Tate's table \cite{Beetal}, already (\ref{weieq}) is a special form of the $SU(3)$ singularity whose parameters are tuned up to ${\cal O}(z^5)$. A change of coordinate $a_1 b_5 +z \to z$ shows the other $SU(2)$ part is also special up to ${\cal O}(z^5)$. The $U(1)_Y$ part is the relative position between two linearly equivalent components. Its global existence depends on the terms in the last line of (\ref{weieq}) although they look sub-leading contribution in $z$, otherwise we cannot have a monodromy-invariant two cycle harboring two-form related to $U(1)_Y$ \cite{CH}.  The Calabi--Yau conditions require that the $b_m$ are sections of $\eta-m c_1$, where $\eta = 6c_1(B_2) + c_1(N_{B_2/B})$ and $c_1 = c_1(B_2)$ are combinations of tangent and normal bundle to $B_2$. The leading order locus of the discriminant in $z$ coincides with $B_2$.

\begin{table}[t]
\begin{center}
\begin{tabular}{ccc} \hline
matter & matter curve & homology on $B_2$ \\ \hline

$q {\bf (3,2;3)}_{\frac16,1,1}$ & $\prod t_i\to 0$    & $\eta-3c_1$  \\
$u^c {\bf (\overline 3,1;3)}_{-\frac23,1,1}$ & $\prod(t_i+t_6)\to 0$ & $\eta-3c_1$  \\
$e^c {\bf (1,1;3)}_{1,1,1} $ & $\prod(t_i-t_6) \to 0$  & $\eta-3c_1$ \\
$d^c {\bf (\overline 3,1;3)}_{\frac13,-3,1} $ & $\prod(t_i+t_5) \to 0$ & $ \eta - 3c_1$  \\
$l {\bf (2,1;3)}_{-\frac12,-3,1}    $  & $\prod(t_i+t_5+t_6) \to 0$
 & $\eta -3c_1$  \\
$\nu^c {\bf (1,1;3)}_{0,5,1}$ & $\prod(t_i-t_5) \to 0$ & $\eta-3c_1$ \\
$h_u^c {\bf (2,1;\overline 3)}_{\frac12,2,2}   $  & $\prod(t_i+t_j+t_6) \to 0$
 & $\eta -3c_1$       \\
$h_d {\bf (2,1;3)}_{\frac12,2,-2} $  & $\prod(t_i+t_4+t_6) \to 0$
 & $\eta -3c_1$ \\
$D_1^c {\bf (\overline 3,1;\overline 3)}_{\frac13,2,2}$ & $\prod(t_i+t_j) \to 0$ & $\eta - 3c_1$ \\
$\bar D_2 {\bf (\overline 3,1;3)}_{\frac13,2,-2} $ & $\prod(t_i+t_4) \to 0$ & $\eta - 3c_1$ \\
$S {\bf (1,1;3)}_{0,0,4}$ & $\prod(t_i-t_4) \to 0$ & $\eta - 3c_1$ \\
\hline
$X{\bf (3,2;1)}_{-\frac56,0,0}$  & $t_6 \to 0$  & $-c_1$  \\
$Y{\bf (3,2;1)}_{\frac16,-4,0}$ & $t_5\to 0$    & $-c_1$   \\
$T^c{\bf (\overline 3,1;1)}_{-\frac23,-4,0} $ & $t_5+t_6 \to 0$ & $-c_1$  \\
$\Sigma{\bf (1,1;1)}_{1,-4,0}$ & $t_5-t_6 \to 0$  & $-c_1$  \\
$Q{\bf (3,2;1)}_{\frac16,1,-3}$ & $t_4\to 0$    & $-c_1$   \\
$U^{c}{\bf (\overline 3,1;1)}_{-\frac23,1,-3} $ & $t_4+t_6 \to 0$ & $-c_1$  \\
$E^c{\bf (1,1;1)}_{1,1,-3}$ & $t_4-t_6 \to 0$  & $-c_1$  \\
$D^{c} {\bf (\overline 3,1;1)}_{\frac13,-3,-3} $ & $t_4+t_5 \to 0$ &  $-c_1$ \\
$L {\bf (2,1;1)}_{-\frac12,-3,-3} $ & $t_4+t_5+t_6$ & $-c_1$ \\
$N^c {\bf (1,1;1)}_{0,5,-3}$ & $t_4-t_5 \to 0$ & $-c_1$ \\
\hline
\end{tabular}
\caption{Matter contents identified by $SU(3)\times SU(2) \times SU(3)_\perp\times U(1)_Y\times U(1)_X \times U(1)_Z$ quantum numbers. All the indices take different value in $S_3=\{1,2,3\}$.
Later, the fields below middle line are decoupled and the charge conjugates of $h_u^c$ and $D_1^c$ will survive as zero modes. }
\label{t:matter}
\end{center} \end{table}

The spectral cover should be further decomposed with smaller structure group, due to phenomenological requirements. 
We need to distinguish Higgs doublets from lepton doublets, having the same SM quantum numbers. The standard way is to introduce the matter parity, or 
its continuous version $U(1)_X$ with the charge being the baryon minus the lepton numbers. This is the commutant to $SU(5)$ inside $SO(10)$ GUT group along $E_n$ series, hence a subgroup of the structure group. So we may decompose the spectral cover with $U(1)_X$. Shortly we will see, for the {\em observed} number of Higgs fields in four dimension, we need one more parameter from an extra $U(1)_Z$, so that the structure group should be factorized as
\begin{equation} \label{factor}
 S[U(3)_\perp \times U(1)_Z \times U(1)_X \times U(1)_Y].
\end{equation}
The resulting spectral cover, respectively $C_3 \cup C_Z \cup C_X \cup C_Y$, is realized by further tuning $d_0 = f_0, d_1=f_1 + e_1 f_0, d_2=f_2+e_1 f_1, d_3 = f_3 + e_1 f_2, d_4= f_4 + e_1 f_3, d_5=e_1 f_4 $ with the constraint $f_0 (b_1+d_1+e_1)+f_1 = 0$.
In $Z$, their classes are respectively $C_3 = 3 \sigma + \pi^* \eta$ and we have linear equivalence relations $C_Z \sim C_X \sim C_Y = \sigma$.

\section{Matter contents}

Since we admit heterotic duality, all four dimensional fields comes from branching and auction of the adjoint $\bf 248$ of $E_8$ \cite{Heckman:2009mn}. Accordingly it branches into multiplets of $SU(3) \times SU(2) \times SU(3)_\perp \times U(1)_Y \times U(1)_X \times U(1)_Z$. The matter spectrum is summarized in Table \ref{t:matter}.
We identify the fields by charge assignments
\begin{equation} \label{u1charges} \begin{split}
 Y&: \textstyle (\frac16,\frac16,\frac16,\frac16,\frac16,-\frac56), \\
 X&: (1,1,1,1,-4,0),\\
 Z&: (1,1,1,-3,0,0),
  \end{split}
\end{equation}
in the basis $\{t_1,t_2,t_3,t_4,t_5,t_6\}$, the weight vectors ${\bf 5}_1+{\bf 1}_{-5}$ of the structure group $S[U(5)\times U(1)_Y]$.
They are localized along curves, the projections of $C_a \cap \tau C_b$ or $C_a \cap C_b,a,b\in \{3,Z,X,Y\}$ on $B_2$, where $\tau$ is involution flipping the orientation of the cover. 

The identities of extra singlets $\nu^c$ and $S$ are understood as follows.  The minimal anomaly free single chiral representation containing all the observed fermions of SM is $\bf 16$ of $SO(10)$. It also contains one extra SM singlet $\nu^c$. Invariance under $SO(10)$ forms Dirac mass term for $\nu^c$ with the SM lepton doublet, thus this is to be interpreted as right-handed neutrino.
With the aid of supersymmetry (SUSY), Higgs bosons belong to a hypermultiplet and can be treated on equal footing as matter. Thus matter and Higgs pair (as well as colored Higgs pair) are unified to a single representation $\bf 27$ of $E_6$. Again it predicts another kind of singlet $S$, and the gauge invariance relates this to $\mu$-parameter of SUSY \cite{Langacker:1998tc}. So the matter contents and couplings naturally show a singlet extension of minimal supersymmetric standard model.

 The field strengths along the Cartan direction come from the dimensional reduction of four-form field strength $G$ of the dual M-theory and this induces vector bundle on the spectral cover \cite{FMW}.
Although the minimal $SU(4)$ $G$-flux preserves unification relation of its commutant $SO(10)$ in $E_8$, the number of Higgs pairs turns out to be completely fixed to be twice the matter multiplicity \cite{CK}. Here we have one more parameter $\zeta$, the trace part of the $U(3)_\perp \subset SU(4)$ vector bundle \cite{Blumenhagen:2005ga}, to relax the condition. So we turn on a universal flux
\begin{equation} \label{fluxes} \textstyle
 \Gamma_3 = \lambda( 3 \sigma - \pi_3^* (\eta - 3 c_1)) + \frac13 \pi_3^*  \zeta,\quad  \Gamma_Z =- \pi_Z^* \zeta, \quad \Gamma_Y = \Gamma_X = 0,
\end{equation}
where $\sigma$ is the class for $B_2$ inside $Z$ and $\pi_3,\pi_Z$ are projections from $U(3)_\perp$ and $U(1)_Z$ covers to $B_2$, respectively. In the F-theory side, we can turn off other fluxes along $U(1)_Y$ or $U(1)_X$ directions, as long as the quantization condition for $\lambda$ below is satisfied. However there is no corresponding picture in the heterotic side, since Fourier-Mukai transformation with zero flux on some of the covers does not make sense.

The number $n_R$ of chiral $R$ zero modes minus anti-chiral $\overline R$ ones of the Dirac operator in $Z$ is a topological number and counted by index theorem. It is simply given by the intersection between matter curve class  and Poincar\'e dual of the $G$-flux, projected on $B_2$ \cite{Donagi:2004ia,Hayashi:2008ba}
\begin{equation} \label{indexthm}
 n_{R} =  {\cal P}_{ R} \cap \Gamma,
\end{equation}
where ${\cal P}_R$ is the matter curve of the representation $R$ and $\cap$ denotes the intersection inside $Z$.
Because of identical geometry of spectral cover as in Ref. \cite{Choi:2011ua,Chen:2010tp}, and we refer to it for the calculation of matter curves
\begin{align}
 \begin{split}
 n_q &= n_{u^c} = n_{e^c} = n_{d^c} = n_l = n_{\nu^c} \\
   &= \textstyle
 (3 \sigma + \eta) \cap \sigma \cap (\lambda(3 \sigma_\infty - \eta)+\frac13 \zeta) + \sigma \cap \sigma \cap (-\zeta)\\
 &  \textstyle = (-\lambda \eta + \frac13 \zeta) \cdot (\eta - 3c_1) +c_1 \cdot \zeta,
 \end{split} \\
 \begin{split}  \label{hu}
 n_{D_1} &= n_{h_u}  \\
 =& \textstyle -(2\sigma + \eta) \cap (\eta - 3c_1) \cap (\lambda(3 \sigma_\infty - \eta)+\frac13 \zeta)  \\
 & \textstyle  = (-\lambda \eta - \frac23 \zeta) \cdot \textstyle (\eta - 3c_1),
\end{split}  \\
\begin{split} \label{hd}
 n_{\bar D_2} &= n_{h_d}  \\
 & =\textstyle (3\sigma + \eta) \cap \sigma \cap  (\lambda(3 \sigma_\infty - \eta)+\frac13 \zeta -\zeta )\\
 & \textstyle  = (-\lambda \eta -\frac23 \zeta) \cdot (\eta-3c_1),
\end{split} \\
 n_X &= n_Y = n_{T^c} = n_{\Sigma} = 0, \\
 n_{Q} &= n_{U^{c}} = n_{E^c} = n_{D^c} = n_{L} = n_{N^c}  =- c_1 \cdot \zeta,
 \label{veclike} \\
\begin{split}
 n_{S}
 & =\textstyle (3\sigma + \eta) \cap \sigma \cap  (\lambda(3 \sigma_\infty - \eta)+\frac13 \zeta +\zeta ) \\
 & = \textstyle ( - \lambda \eta+\frac43 \zeta) \cdot (\eta - 3c_1) .
\end{split}
\end{align}
Here we omitted pullback and the dot product is for the divisors of $B_2$. We defined $\sigma_\infty = \sigma + \pi^* c_1$.
All the matter fields appearing here are those inside $\bf 27$ multiplet of $E_6$. Their multiplicities manifest the $SO(10)$ unification relation, predicting the same number of right-handed neutrinos. They are preserved because the $G$-flux is along $SU(3)_\perp \times U(1)_Z$ structure group. It is a nontrivial check that $h_u$ and $h_d$ gives the same number in (\ref{hu}) and (\ref{hd}).

The numbers of matter generations and Higgs pairs can  be {\em individually} controlled, depending on the topological data on $B_2$. We require three generations of matter and one pair of Higgs doublets
\begin{equation} \textstyle
 \lambda \eta \cdot (\eta-3c_1) = -\frac73, \quad \eta \cdot \zeta =2, \quad c_1 \cdot \zeta = 0.
\end{equation}
They are subject to quantization conditions $3(\frac12 + \lambda) \in \Z, (\frac12-\lambda)\eta+(3\lambda-\frac12)c_1+\frac13 \zeta \in H_2(S,\Z)$ where $\lambda$ is a nonnegative rational number. We find a solution $\lambda=\frac16$, for which only an integral or half-integral $\lambda$ is possible in the absence of $U(1)_Z$ flux $\zeta$. The base as del Pezzo two surface with  $\eta = 2H, \zeta = H -3 E_1$ do the job, where $H$ is hyperplane divisor and $E_1$ is one of the exceptional divisor. 
This relation restricts the number of the SM neutral field $S$ be five.
In addition, because $a_1$ in (\ref{su5u1cover}) transforms as a section of $-c_1$, we have two scalar fields $O$ and $O'$ transforming as adjoints under $S[U(3)\times U(2)] \simeq SU(3)\times SU(2)\times U(1)$, belonging to $H^{2,0}(B_2)+H^{0,1}(B_2)$ \cite{DW3}. They will play an interesting role in vacuum configuration around the string scale $M_s$.
The other $E_8$ serves as hidden sector and is completely decoupled in smooth compactification and it can serve as supersymmetry breaking sector.
In the F-theory side, we can turn off other fluxes along $U(1)_Y$ or $U(1)_X$ directions, as long as the quantization condition for $\lambda$ is satisfied.

\section{Higgs sector and nucleon decay}

The requirement of {\em one pair} of Higgs doublets fixed the factorization of spectral cover (\ref{factor}). It has the following phenomenological implications.

Firstly, it also distinguishes between {\em up and down} type Higgses. This is due to the structure of the $SU(3)_\perp$ monodromy $S_3$ \cite{Tatar:2009jk}, the permutation of the elements $\{t_1,t_2,t_3\}$. It is the natural Weyl group, without a special monodromy further selected by hand. In terms of the $S_3$ representations, the fields having the same quantum number of lepton doublet under the SM group are
\begin{equation} \label{leptons} \begin{split}
 l&: \{t_1+t_5+t_6,t_2+t_5+t_6,t_3+t_5+t_6\}, \\
 h^c_u&: \{t_1+t_2+t_6,t_2+t_3+t_6,t_3+t_1+t_6\}, \\
 h_d&: \{ t_1+t_4+t_6,t_2+t_4+t_6,t_3+t_4+t_6\}, \\
 L&: \{t_1+t_5+t_6,t_2+t_5+t_6,t_3+t_5+t_6\}.
\end{split}
\end{equation}
Effectively, the Higgs doublet is distinguished from the lepton doublet by an opposite matter parity or the $U(1)_X$. It also forbids bare (super)renormalizable lepton and/or baryon number violating operators $lh_u,lle^c,lqd^c,u^c d^c d^c$.
Further factorization ruins this one Higgs pair structure but we obtain three pairs of Higgses, so our factorization seems the unique for the $U(n)$ type spectral cover with universal flux.

Well-known is that the matter parity and $U(1)_X$ alone cannot forbid dimension five proton decay operators such as $qqql$ and $u^c u^c d^c e^c$. However, the above structure group {\em forbids} these operators. For instance, $qqql$ is not allowed because of nonvanishing sum of the weights $(t_i)+(t_j)+(t_k)+(t_i+t_5+t_6)$ and $u^c u^c d^c e^c$ is not because of $(t_i+t_6)+(t_j+t_6)+(t_k+t_5)+(t_i-t_6)$, required by $SU(3)_\perp$ invariance, since one of $S_3$ index should appear twice \cite{pdecay}. At the field theory level, this is also simply understood by invariance under $U(1)_Z$ \cite{U1Z}

Another prediction is the presence of an SM singlet field $S$. Surveying the quantum number, it belongs to $\bf 27$ representation of $E_6$, therefore, its interaction is restricted and we can calculate the corresponding terms.
Because Higgs doublet and triplets are not simply vectorlike and up and down Higgses live on different matter curves, bare masses are forbidden by $SU(3)_\perp$ invariance. Instead we have a singlet extension to MSSM \cite{NMSSM,Cvetic:2010dz}. We can check that the only renormalizable superpotential for the surviving fields are (see also below)
\begin{equation}  \label{pertW} \begin{split}
 q h_d u^c &+ q h_u d^c + l h_d e^c +l h_u \nu^c + S h_u h_d +S D_1 \bar D_2 \\
 +&\ q q D_1 + u^c e^c D_1 + q l \bar D_2 + \nu^c d^c D_1+u^c d^c \bar D_2
  \end{split} \end{equation}
omitting the flavor dependent coefficients. We expect the terms involving $D_1$ and $D_c^2$ are all decoupled, yielding the $\mu$-like term $S h_u h_d$. Bare quadratic or cubic terms in $S$ are not allowed by invariance under the $SU(3)_\perp$ and other $U(1)$'s. Induced higher order terms include $
 M_{s}^{-2} SS ( Q \bar D_2 L  + D^cN D_1  + U^c E^c D_1)  + M_{s}^{-4} SSS (  Q U^c E^c L + Q D^c N^c L )$ but they are to be suppressed by a string scale $M_s$.
A Majorana mass for the $\nu^c$ does not appear up to dimension five. There is an interesting room for this from Euclidean D3-brane or gauge instanton in F-theory \cite{Finst}, which might as well generate similar potential for $S$. 

Since the Higgs fields also obey $SO(10)$ unification relation, we have as many colored Higgs pairs $D_1,\bar D_2$ as doublets. This doublet-triplet splitting problem should be solved by an effect evading the unification structure, close but below $M_s$.
It is a possibility to consider a {\em vectorlike} extra generation of matter fields, without changing the Dirac indices.
Using aforementioned $U(3)$ adjoint chiral super field $O$, there can be terms $\langle O \rangle D_1 D_1^c + \langle O \rangle \bar D_2 \bar D_2^c + M_O \tr O^2 + \tr O^3$ giving Dirac masses separately to $D_1,D_1^c$ and $\bar D_2,\bar D_2^c$ pairs. Conventional gauge coupling unification requires heavy triplets, so do a large VEV $\langle O \rangle$ and a large $M_O$ \cite{Perez:2008ry}. We can allow also vectorlike pair for the doublet, but in principle a similar $U(2)$ adjoint can give different masses. This seems like a flavor problem in the UV regime and more is to be understood.
On the other hand,
we expect a coupling $(\langle S \rangle + \mu_D) D_1 \bar D_2$ is generated, with a possible SUSY breaking effect $\mu_D$ \cite{mu}. The most strongly constrained nucleon decay operator is $qqql$, whose coefficient has upper bound $10^{-5} M_P^{-1}$ \cite{Raby:2002wc}. At low energy scale, integrating out heavy fields, $qqD_1$ and $ql \bar D_2$ may induce an operator $(\langle S \rangle + \mu_D)/ M_D^{-2} qqql$ up to geometric suppression factor. Once forbidden at the tree-level, it is also known that the induced operators are highly suppressed, probably explained by worldsheet instanton contribution \cite{wsinst}.  A possible mixing from bare mass term $d^c d$ does not change this eigenvalue. The same argument goes to other induced operators for nucleon decay.

\section{Anomalous $U(1)$}

We check the $G$-flux contribution to $D$-term for each $U(1)$ using type IIB string limit \cite{Grimm:2011tb}, where we have Ramond--Ramond four-form field $C_4$ in low energy. Its Kaluza--Klein expansion along a harmonic two-form $w_2 \in H^{1,1}(B_2, \Z)$ has a form $C_4 = C_2 \wedge \omega_2$, yielding the interaction
$
 {\rm tr}\, t_Q^2 \int_{M^4} F_Q \wedge C_2 \int_{B_2} i^* \omega_2  \wedge  \langle F_Q \rangle 
$
from Chern--Simons interactions and here $F_Q$, generated along $t_Q$ direction, is the field strength for $U(1)_Q$ flux and $i$ is immersion to $B$.
We turned on a flux for $U(1)_Z$ as in (\ref{fluxes}) thus the corresponding gauge boson acquire mass by St\"uckelberg mechanism and the symmetry is broken. On the heterotic side, it looks that the anomaly of $U(1)_Z$ is removed by shift of model-dependent axion \cite{DG}, which is the imaginary part of superfield $T = \int_Q J + i \int_Q B$, where $J$ is the K\"ahler form, $B$ is the NSNS two-form, and $Q$ is interpreted as two-cycle wrapped by worldsheet instanton \cite{MSW}. Now $T$ is charged and there is an instanton generating a nonperturbative super potential, guided by $U(1)_Z$ invariance.

To keep $SO(10)$ unification relation for the matter multiplicity, we do not turn on flux along $X$ direction, and the only possible superpotential is of a form $e^{-T} S^n, n \in \Z$. In this case, $U(1)_X$ and hypercharge do not belong to the structure group of the vector bundle in the heterotic side, and they may remain as unbroken group in the low energy \cite{DG}. Phenomenology of these extra $U(1)$ groups inside $E_6$ are recently discussed in Ref. \cite{Erler:2011iw}. 

Since we do not turn touch other unbroken gauge group, their gauge couplings receive no threshold correction from the flux from F-theory side \cite{Blumenhagen:2008aw,Donagi:2008kj}. The four dimensional gauge coupling is inversely proportional to the volume of four cycle $S$ supporting gauge group, but to be precise it is topologically given by intersection numbers $g^{-1}_{\rm 4D} \propto e^{-\phi} \int_S J \wedge J $. Since $SU(3)$ and $SU(2)$ have linearly equivalent cycle \cite{Choi:2010nf,Choi}, we have the same four dimensional coupling. In fact, we have only one gauge coupling of embedded in $E_8$, and $SO(10)$, giving the same coupling to $SU(3) \times SU(2) \times U(1)_Y \times U(1)_X$ with the correct normalization in $SO(10)$
$$
  g_3 = g_2 = \sqrt{\frac53} g_Y = \sqrt{40} g_X, \ \sin^2 \theta_W = \frac38,
$$
at $M_s$. 
The $U(1)_X$ can survive as gauge symmetry at relatively low energy scale and would be spontaneously broken down at relative low energy.
Threshold corrections for the splitted Higgs triplets $D_1$ and $\bar D_2$ would modify the scale.

\section{Conclusion}
We sought a supersymmetric extension of the Standard Model using spectral cover construction. As a minimal set of conditions, we required the SM group, matter parity, and the correct number of the Higgs doublet. Each step narrowed the structure group of the vector bundle to a subgroup of $S[U(5)\times U(1)], S[U(4)\times U(1)^2], S[U(3)_\perp \times U(1)^3]$, respectively. Since a smaller structure group such as $S[U(2)\times U(1)^4]$ cannot reproduce the desired spectrum and couplings, the only possible choice in this framework is $S[U(3)_\perp \times U(1)^3]$.  Requiring three generations of matter fields and one pair of Higgs doublets, the universal $G$-flux is turned on with the structure group $S[U(3)_\perp \times U(1)_Z]$, resulting in the multiplicity of the spectrum satisfying $SO(10)$ unification relation. Another flux component along $U(1)_X$ is optional. As a nontrivial consequence of the spectral cover and the resulting matter localization, we are able to distinguish up and down Higgs fields, and obtain a restricted perturvative and nonperturbative superpotentials for the singlets $S$ giving $\mu$-term. Analysis of the consequent dynamics would be an interesting future direction.

\begin{acknowledgements}
The author is grateful to Hirotaka Hayashi, Jihn E. Kim, Tatsuo Kobayashi, Bumseok Kyae, Seung-Joo Lee, Hans-Peter Nilles and Martijn Wijnholt for discussions and correspondences. This work is partly supported by the National Research Foundation of Korea (NRF) with grant number 2012-040695.
\end{acknowledgements}

\end{document}